\documentclass[prd,aps,preprint,nofootinbib,superscriptaddress]{revtex4}
\usepackage{epsfig}
\usepackage{amsmath}
\input{epsf}

\begin{document}


\vspace*{2cm}
\title{A note on the scale evolution of the ETQS function $T_F(x,x)$}

\author{Andreas Sch\"{a}fer}
\affiliation{\normalsize\it Institut f\"{u}r Theoretische
Physik,Universit\"{a}t Regensburg, Regensburg, Germany}
\author{Jian Zhou}
\affiliation{\normalsize\it Institut f\"{u}r Theoretische
Physik,Universit\"{a}t Regensburg, Regensburg, Germany}

\begin{abstract}
We reexamine the scale dependence  of the ETQS
(Efremov-Teryaev-Qiu-Sterman) twist-3 matrix element which has been
studied already by the four different groups with conflicting
results~\cite{Kang:2008ey,Zhou:2008mz,Vogelsang:2009pj,Braun:2009mi}.
We find that we can in fact reproduce the results of
~\cite{Braun:2009mi} with the method~\cite{Zhou:2008mz} when we
treat some subtleties with greater care, thus easing the mentioned
conflict.
\end{abstract}

\maketitle The ETQS matrix element plays an important role for the
theoretical description of transverse single spin asymmetries (SSA)
in the framework of collinear factorization. The control of
$Q²$-evolution is not only necessary to describe QCD dynamics
correctly and to reduce the dependence of theory predictions on the
factorization scale adopted, but such evolution equations give also
insight into the functional form of higher-twist distribution
functions. The idea there is to start evolution at a low scale and
use the fact that the resulting form at a high scale is only little
dependent on the low-scale input~\cite{Braun:2011aw}. The latter is
especially important in view of the limited experimental input of
has to determine these functions.

The corresponding calculation was done in
Refs.~\cite{Kang:2008ey,Zhou:2008mz,Vogelsang:2009pj,Braun:2009mi}.
However, the result obtained in Ref.~\cite{Braun:2009mi}  differ
from that derived in
Refs.~\cite{Kang:2008ey,Zhou:2008mz,Vogelsang:2009pj} by two extra
terms. It was settled rather easily that one of these terms is due
to a Feynman diagram which was missed in
Refs.~\cite{Kang:2008ey,Zhou:2008mz,Vogelsang:2009pj}. The second
additional term in~\cite{Braun:2009mi} which is proportional to
$\delta(1-z)$ could not be reproduced by the other calculations so
far. We show in this short contribution, how this term arises within
the formalism of Ref.~\cite{Zhou:2008mz} due to a rather subtle fact
related to the non-commutativity of a certain limit and a certain
integration. We now hope to be able to do this calculation
consistently in the light cone gauge.

The ETQS function $T_F$ is defined through the following matrix
element,
\begin{eqnarray}
 \int \frac{dy^-}{2 \pi}
\frac{dy_1^-}{2\pi} e^{ixP^+ y^-}  \langle PS | \bar{\psi}_\beta(0)
g F^{+\mu}(y_1^-)\psi_\alpha(y^-) | PS \rangle  \ = \frac{M}{2}
T_F(x,x)\epsilon^{\nu\mu}_\perp S_{\perp \nu} p\!\!\!/
\end{eqnarray}
In Ref~\cite{Zhou:2008mz},  the light-cone gauge ($A^+=0$) with the
retarded boundary condition, i.e., $A_\perp(-\infty^-)=0$ was chosen
such that $T_F(x,x)$ can be rewritten as,
\begin{eqnarray}
T_F(x)=\int \frac{dy^-}{8 \pi^2 M} e^{ixP^+ y^-} \langle PS |
\bar{\psi}(0) n\!\!\!/ \epsilon^{\nu\mu}_\perp S_{\perp \nu}
i\partial_{\perp\mu} \psi_\alpha(y^-) | PS \rangle \ .
\end{eqnarray}

To calculate the splitting function, one has to take into account
the contributions from the operators $\left(\bar\psi\partial_\perp
\psi\right)$ and $\left(\bar\psi A_\perp\psi\right)$, because they
are of the same twist. We plot the Feynman diagrams contributions
for the real gluon radiation in Fig.~1, where (a) is the
contribution from the partial derivative acting on the quark field,
and $(b-d)$ are those from $A_\perp$ contributions. Virtual
corrections only contribute to the contribution proportional to
$\left(\bar\psi\partial_\perp \psi\right)$. Their contribution is
the same as for the quark self energy correction.

\begin{figure}[t]
\begin{center}
\includegraphics[width=10cm]{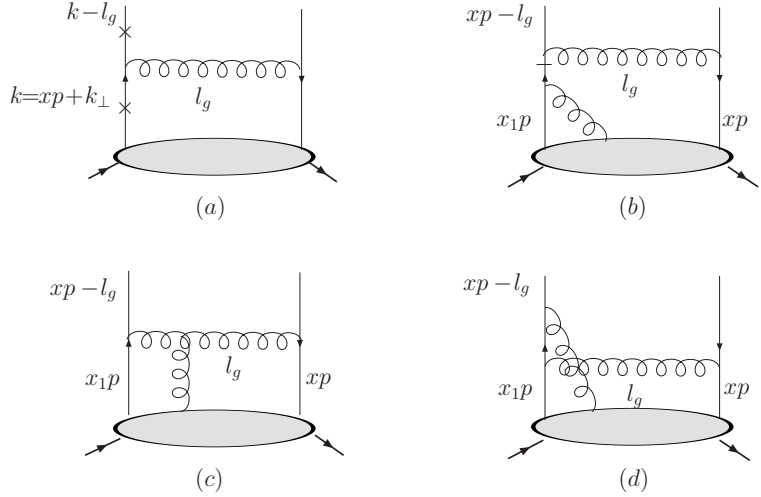}
\end{center}
\vskip -0.4cm \caption{Real gluon radiation contribution to the
evolution equation for the ETQS function $T_F(x,x)$. Crosses in
fig.(a) and horizontal bar in fig.(b) indicate $k_\perp$ flow and
special propagator, respectively. } \label{fig1}
\end{figure}

Following the procedure presented in Ref.~\cite{Zhou:2008mz}, we
perform a collinear expansion for the hard scattering part to
calculate the contribution from Fig.~1(a). The linear $k_\perp$
expansion term combining with the quark field will lead to the
quark-gluon correlation function $T_F(x,x)$. In the collinear
expansion in terms of $k_\perp$, we can fix the transverse momentum
of the probing quark ($l_q$) or the radiated gluon ($l_g$), because
of momentum conservation and we are integrating over them to obtain
$T_F(x,x)$. We have also checked that they will generate the same
result. Following the Ref.~\cite{Zhou:2008mz}, we fix $l_g$ in the
collinear expansion to simplify the calculation.

For the $A_\perp$ contribution, we notice that $F^{+\mu}=\partial^+
A_\perp^\mu$ in the light cone gauge. Therefore, one can relate the
corresponding soft matrix to the correlation function $T_F(x,x_1)$
in the following way,
\begin{eqnarray}
&& \frac{i}{x-x_1+i\epsilon}\int \frac{dy^-dy_1^-}{4 \pi} e^{ix_1P^+
y^-} e^{i(x-x_1)P^+y_1^-}\langle PS | \bar{\psi}_\beta(0^-) n\!\!\!/
\epsilon^{\nu\mu}_\perp S_{\perp \nu} g{F^+}_\mu(y_1^-)
\psi_\alpha(y^-) | PS \rangle
\nonumber \\
&=&\int \frac{dy^-dy_1^-}{4 \pi }P^+ e^{ix_1P^+ y^-}
e^{i(x-x_1)P^+y_1^-}\langle PS | \bar{\psi}_\beta(0^-) n\!\!\!/
\epsilon^{\nu\mu}_\perp S_{\perp \nu} gA_{\perp\mu}(y_1^-)
\psi_\alpha(y^-) | PS \rangle \ .
\end{eqnarray}
In above formula, the soft gluon pole appearing in the first line is
generated by partial integration. The pole prescription has been
determined by our choice of a retarded boundary condition. For the
same reason, we have to regulate the light cone propagator in a
consistent manner.  The gluon propagator appearing in Fig.~1(c) in
the light cone gauge with the retarded boundary condition is given
by,
\begin{eqnarray}
D^{\alpha\beta}(l)=\frac{-i}{l^2+i\epsilon} \left ( g^{\alpha\beta}
- \frac{l^\alpha n^\beta +n^\alpha l^\beta}{l \cdot n + i\epsilon}
\right ) \ ,
\end{eqnarray}
where $l$ is the gluon propagator momentum flowing out from the
quark-gluon vertex in Fig.1(c).

We now deviate from the original calculation \cite{Zhou:2008mz} in
two ways:

i) in~\cite{Zhou:2008mz} the integral $\int_{x_g'}   d x_g
\frac{x_g' \delta(x_g')}{x_g^2}$ was simply neglected;

ii)  one of the two absorptive parts of the free propagator was not
taken into account.

We  will discuss next these two points  in more details, arguing
that the neglected contributions have to be taken into account. When
computing the hard pole contribution from Fig.1(c), for the left cut
diagram, one has
\begin{eqnarray}
T_F^{(1)}|_{\rm Fig.1(c)}^{\rm hp-left}(x_B)&=&\frac{\alpha_s}{4\pi}
\int_{x_B} \frac{dx }{x}\int_{x_g'}dx_g
\frac{dl_{g\perp}^2}{l_{g\perp}^2} \frac{C_A}{2} \left [
\frac{(x+x_B)(2x_g-x_g')}{2 x_g^2} \right ] \delta(x_g')T_F(x-x_g,x)
\
\end{eqnarray}
where $x_g'=l \! \cdot \!  n/p^+$ with $x_g'=x_B-x+x_g$. By noticing
that $\int_{x_g'}   d x_g \frac{x_g'
\delta(x_g')}{x_g^2}=\delta(x_B-x)$, rather than zero, and summing
left and right cut diagrams, one obtains,
\begin{eqnarray}
T_F^{(1)}|_{\rm Fig.1(c)}^{\rm hp}(x_B)&=&\frac{\alpha_s}{2\pi}
\int_{x_B} \frac{dx }{x} \frac{dl_{g\perp}^2}{l_{g\perp}^2}
\frac{C_A}{2} \left [
\frac{1+z}{1-z}T_F(x_B,x)-\delta(1-z)T_F(x_B,x_B) \right ] \
\end{eqnarray}
where $z=x_B/x$. The second term is missing in
Ref.\cite{Zhou:2008mz}.

Next we discuss the second contribution which was overlooked in
Ref.\cite{Zhou:2008mz}. Since we work in the light-cone gauge with
retarded boundary condition, the free propagator possesses two
absorptive parts~\cite{Bassetto:1996ph},
\begin{eqnarray}
discD^{\alpha\beta}(l_g)=2\pi \theta(l_g^0) \delta(l_g^2) \left [
-g^{\alpha\beta} + \frac{2l_g^-(l_g^\alpha n^\beta +n^\alpha
l_g^\beta)}{l_{g\perp}^2} \right ]-2 \pi \theta(l_g^0) \delta(l_g^+)
\frac{(l_g^\alpha n^\beta +n^\alpha l_g^\beta)}{l_{g\perp}^2}
\end{eqnarray}
In \cite{Zhou:2008mz} only the first absorptive part was taken into
account. In order to carry out the calculation in a consistent
manner, one must include the contribution from the second part.
However, if one still picks up the same imaginary part as we did
above, this contribution will cancel between the different cut
diagrams as it happens when both gluon lines go on shell. On the
other side, the additional imaginary part may come from the
artificial pole which appears in Eq.(3).  Such pole-absorptive part
combination gives the contribution,
\begin{eqnarray}
T_F^{(1)}|_{\rm Fig.1(c)}^{\rm LC-left}(x_B)&=&
\frac{\alpha_s}{4\pi} \int \frac{dl_{g\perp}^2}{l_{g\perp}^2}
\int_{x_B} dx  \int_0^\infty dl_g^- \int_{x_g'}dx_g
\delta(x_g-x_g') \delta(x_g) \nonumber \\ && \times \frac{C_A}{2}
\left [
\frac{2(2x_B-x_g')(x_g+x_g')}{2(2l_g^-x_B+l_{g\perp}^2)x_g'}-
 \frac{2x_B l_{g\perp}^2}{(2l_g^-x_B+l_{g\perp}^2)^2}\right ] T_F(x-x_g,x) \
\end{eqnarray}
Integrating over $x_g, l_g^-$ and summing the two cut diagrams, we
obtain,
\begin{eqnarray}
T_F^{(1)}|_{\rm Fig.1(c)}^{\rm LC}(x_B)&=& \frac{\alpha_s}{2\pi}
\int \frac{dl_{g\perp}^2}{l_{g\perp}^2}  \int_{x_B} \frac{dx }{x}
\frac{C_A}{2} \left [ \int_0^1 dy \frac{2}{1-y}-
 1\right ] \delta(1-z) T_F(x_B,x_B) \
\end{eqnarray}
Taking into account the contribution from the second part of the
absorptive part, Eq.(20) in the Ref.\cite{Zhou:2008mz} should be
modified as follows,
\begin{eqnarray}
&&-\frac{\alpha_s}{2\pi} \frac{C_A}{2} \int_{x_B}
\frac{dx}{x}T_F(x,x)  d^2l_{g\perp} \left [ \frac{
\partial}{\partial l_{g\perp}^\mu } \hat H_0(xP,l_{g\perp} ) \right
] \times (-l_{g\perp}^\mu)
\nonumber \\
&=&-\frac{\alpha_s}{2\pi}  \frac{C_A}{2} \int_{x_B} \frac{dx}{x}
T_F(x,x)  d^2l_{g\perp}
\hat H_0(xP,l_{g\perp} )\nonumber\\
&=&-\frac{\alpha_s}{2\pi} \frac{C_A}{2} \int_{x_B} \frac{dx}{x}
\frac{dl_{g\perp}^2}{l_{g\perp}^2}\left[
\frac{1+z^2}{1-z}-\delta(1-z) \int_0^1 dy \frac{2}{1-y} \right ]
T_F(x,x) \ .
\end{eqnarray}
Collecting all pieces, we eventually arrive at the following scale
evolution equation for $T_F(x,x)$,
\begin{eqnarray}
&&\!\!\!\!\!\!\!\!\!\!\!\!\!\!\!\!\! \frac{\partial
T_F(x_B,x_B,\mu^2)}{\partial {\ln\mu^2}} =\frac{\alpha_s}{2\pi}
\int_{x_B} \frac{dx}{x} \left [ C_F \left \{ \frac{1+z^2}{(1-z)_+}+
\frac{3}{2} \delta(1-z) \right \} T_F(x,x) \right .\
\nonumber \\
&& \ \ \ + \left .\ \!\!\! \frac{C_A}{2} \left \{ \frac{1+z}{1-z }
T_F(xz,x) -\frac{1+z^2}{1-z}  T_F(x,x)-2\delta(1-z)T_F(x,x)+
\tilde{T}_F(xz,x) \right \} \right ] \ ,
\end{eqnarray}
which coincides with the result given in Ref.~\cite{Braun:2009mi}.

As shown above, the missing boundary term $-2\delta(1-z)T_F(x,x)$
appears to be a generic problem which might have far reaching
consequences. In principle, all previous calculations involving hard
gluon pole contributions might need to be reexamined. However, one
can expect that the observed matching between the TMD factorization
and collinear factorization at intermediate transverse momentum will
not be affected by this extra term.

\noindent {\bf Acknowledgments:} When this paper was finishing, we
learned that the extra boundary term also can be recovered in both
Kang-Qiu's approach and Vogelsang-Yuan's approach~\cite{VY, Kang}.
This work has been supported by BMBF (OR 06RY9191). We thank
Alexander Manashov and Vladimir Braun for a lot of very useful
discussions and encouragement.

\end {document}